\documentclass[aps,twocolumn,showpacs,preprintnumbers,superscriptaddress]{revtex4}
\usepackage{amsmath}
\usepackage{amssymb}
\usepackage{mathrsfs}
\usepackage{graphicx}
\usepackage{longtable}
\usepackage{tabularx}
\usepackage{float}
\usepackage{bm}
\usepackage{natbib}
\usepackage{braket}
\usepackage{epstopdf}
\usepackage[colorlinks=true,citecolor=blue,linkcolor=blue,urlcolor=blue]{hyperref}

\begin{document}

\title{Electric-field modulation of linear dichroism and Faraday rotation in few-layer phosphorene}

\author{L. L. Li}
\email{longlong.li@uantwerpen.be}
\address{Department of Physics, University of Antwerp,
Groenenborgerlaan 171, 2020 Antwerpen, Belgium}

\author{B. Partoens}
\address{Department of Physics, University of Antwerp,
Groenenborgerlaan 171, 2020 Antwerpen, Belgium}

\author{W. Xu}
\address{Key Laboratory of Materials Physics, Institute of
Solid State Physics, Chinese Academy of Sciences, Hefei 230031,
China}
\address{School of Physics and Astronomy and Yunnan Key Laboratory for Quantum Information, Yunnan University, Kunming 650091, China}

\author{F. M. Peeters}
\email{francois.peeters@uantwerpen.be}
\address{School of Physics and Astronomy and Yunnan Key Laboratory for Quantum Information, Yunnan University, Kunming 650091, China}
\address{Department of Physics, University of Antwerp,
Groenenborgerlaan 171, 2020 Antwerpen, Belgium}

\date{\today}

\begin{abstract}
Electro-optical modulators, which use an electric voltage (or an electric field) to modulate a beam of light, are essential elements in present-day telecommunication devices. Using a self-consistent tight-binding approach combined with the standard Kubo formula, we show that the optical conductivity and the linear dichroism of few-layer phosphorene can be modulated by a perpendicular electric field. We find that the field-induced charge screening plays a significant role in modulating the optical conductivity and the linear dichroism. Distinct absorption peaks are induced in the conductivity spectrum due to the strong quantum confinement along the out-of-plane direction and to the field-induced forbidden-to-allowed transitions. The field modulation of the linear dichroism becomes more pronounced with increasing number of phosphorene layers. We also show that the Faraday rotation is present in few-layer phosphorene even in the absence of an external magnetic field. This optical Hall effect is induced by the reduced lattice symmetry of few-layer phosphorene. The Faraday rotation is greatly influenced by the field-induced charge screening and is strongly dependent on the strength of perpendicular electric field and on the number of phosphorene layers.
\end{abstract}

\maketitle

\section{Introduction}

Electro-optical modulators use an electric voltage (or an electric field) to modulate a beam of light. The modulation may be imposed on the phase, frequency, amplitude, or polarization of the light beam. The operation principle is based on the Franz-Keldysh effect \cite{williamsElectricFieldInduced1960} where the band-gap energy and the optical absorption of a bulk material can be changed by an applied electric field. In a quantum-well configuration, the energy of the quantized subbands can also be changed by an external electric field allowing the modulation of the absorption spectrum at the different onset energies of the subbands. Anisotropic materials (such as wurtzite semiconductors) exhibit optical dichroism \cite{shokhovetsAnisotropyMomentumMatrix2008} in which light having different polarizations are absorbed by different amounts. Here we show that in few-layer phosphorene a perpendicular electric field can modulate the optical conductivity and the linear dichroism, which makes it a promising new material for optoelectronic applications in e.g. electro-optical modulators.

Since its first isolation in 2014 \cite{liuPhosphoreneUnexplored2D2014a}, phosphorene, a single layer of black phosphorus (BP), has drawn a lot of attention from the research community due to its extraordinary properties \cite{kouPhosphoreneFabricationProperties2015, carvalhoPhosphoreneTheoryApplications2016}. This new two-dimensional (2D) material features a direct band gap combined with a high carrier mobility \cite{kouPhosphoreneFabricationProperties2015, carvalhoPhosphoreneTheoryApplications2016}, which is crucial for practical applications in e.g. field-effect transistors. Another striking feature is that due to the puckered lattice structure phosphorene exhibits strongly anisotropic electronic, optical, transport, and plasmonic properties such as strong anisotropy in the energy spectrum, effective mass, optical absorption, electrical mobility, and plasmonic excitation \cite{qiaoHighmobilityTransportAnisotropy2014d, tranLayercontrolledBandGap2014b,cakirTuningElectronicOptical2014a, yuanTransportOpticalProperties2015c,lowPlasmonsScreeningMonolayer2014b}.

Multilayer phosphorene is composed of two or more phosphorene monolayers that are stacked and coupled via the van der Waals interaction. The band gap, which remains direct in multilayer phosphorene, decreases monotonically with increasing number of stacking layers (from $\sim$ 2 eV for monolayer down to $\sim$ 0.3 eV in the bulk limit) \cite{qiaoHighmobilityTransportAnisotropy2014d, tranLayercontrolledBandGap2014b}. It was predicted that by applying a perpendicular electric field, the band gap of multilayer phosphorene can be closed and as a result, a phase transition from a normal semiconductor to a Dirac semimetal can be induced in multilayer phosphorene \cite{liuSwitchingNormalInsulator2015a, pereiraLandauLevelsSinglelayer2015b,yuanQuantumHallEffect2016c, wuFieldinducedDiverseQuantizations2017b}, leading to the appearance of a linear energy spectrum and a zeroth Landau level \cite{yuanQuantumHallEffect2016c, wuFieldinducedDiverseQuantizations2017b}.

Most recently, we showed \cite{liTuningElectronicProperties2018c} that by applying a perpendicular electric field, both intralayer and interlayer charge screening can be induced in multilayer phosphorene and the field-induced charge screening plays a significant role in tuning the electronic properties of multilayer phosphorene. We found \cite{liTuningElectronicProperties2018c} that it was essential to include the field-induced charge screening in order to achieve good agreement between theory and experiment for the electric-field tuning of the band gap of multilayer phosphorene. Such electric-field tuning of the electronic properties will result in a modification of the optical properties. The optical properties of multilayer phosphorene were theoretically investigated by means of the effective $\textbf{k}\cdot\textbf{p}$ approach \cite{lowTunableOpticalProperties2014c,linMultilayerBlackPhosphorus2016a}, where the dependence of the optical conductivity on the layer thickness, the carrier doping, and on the magnetic field were obtained. However, this continuum approach is no longer applicable for multilayer phosphorene with fewer stacking layers where the distinct atomic nature of layered structure becomes important. Furthermore, the field-induced charge screening that was shown to be important in few-layer phosphorene (e.g., in bilayer and trilayer phosphorene) \cite{liTuningElectronicProperties2018c} was also not included in those studies \cite{lowTunableOpticalProperties2014c,linMultilayerBlackPhosphorus2016a}. Recent experiments reported electrically tunable linear dichroism in few-layer phosphorene \cite{pengMidinfraredElectroopticModulation2017c, sherrottElectricalControlLinear2017a}, where the active modulation of the linear dichroism by an external gate voltage was measured from the infrared to the visible frequency range.

In the present work, we investigate theoretically the modulation of the optical properties of few-layer phosphorene by a perpendicular electric field. We use a self-consistent tight-binding (TB) approach, which takes account of the field-induced charge screening, to calculate the energy spectrum and the wave function of few-layer phosphorene. Based on the self-consistent TB results, we use the Kubo formula to calculate the interband optical conductivity of few-layer phosphorene over a wide range of photon energies covering the infrared-to-visible frequency range. We show that the interband optical conductivity in the infrared-to-visible frequency range is strongly dependent on the perpendicular electric field and is greatly influenced by the field-induced charge screening. Particular attention is paid to the electric-field modulation of the linear dichroism and the Faraday rotation in few-layer phosphorene. We find that the linear dichroism and the Faraday rotation in few-layer phosphorene can strongly be modulated by the perpendicular electric field, and that the field-induced charge screening plays a significant role in the modulation. Our theoretical study is relevant for electrically tunable linear dichroism and Faraday rotation in few-layer phosphorene in the infrared-to-visible frequency range.

\section{Theoretical Approach}

\begin{figure}
\center
\includegraphics[width=0.49\textwidth]{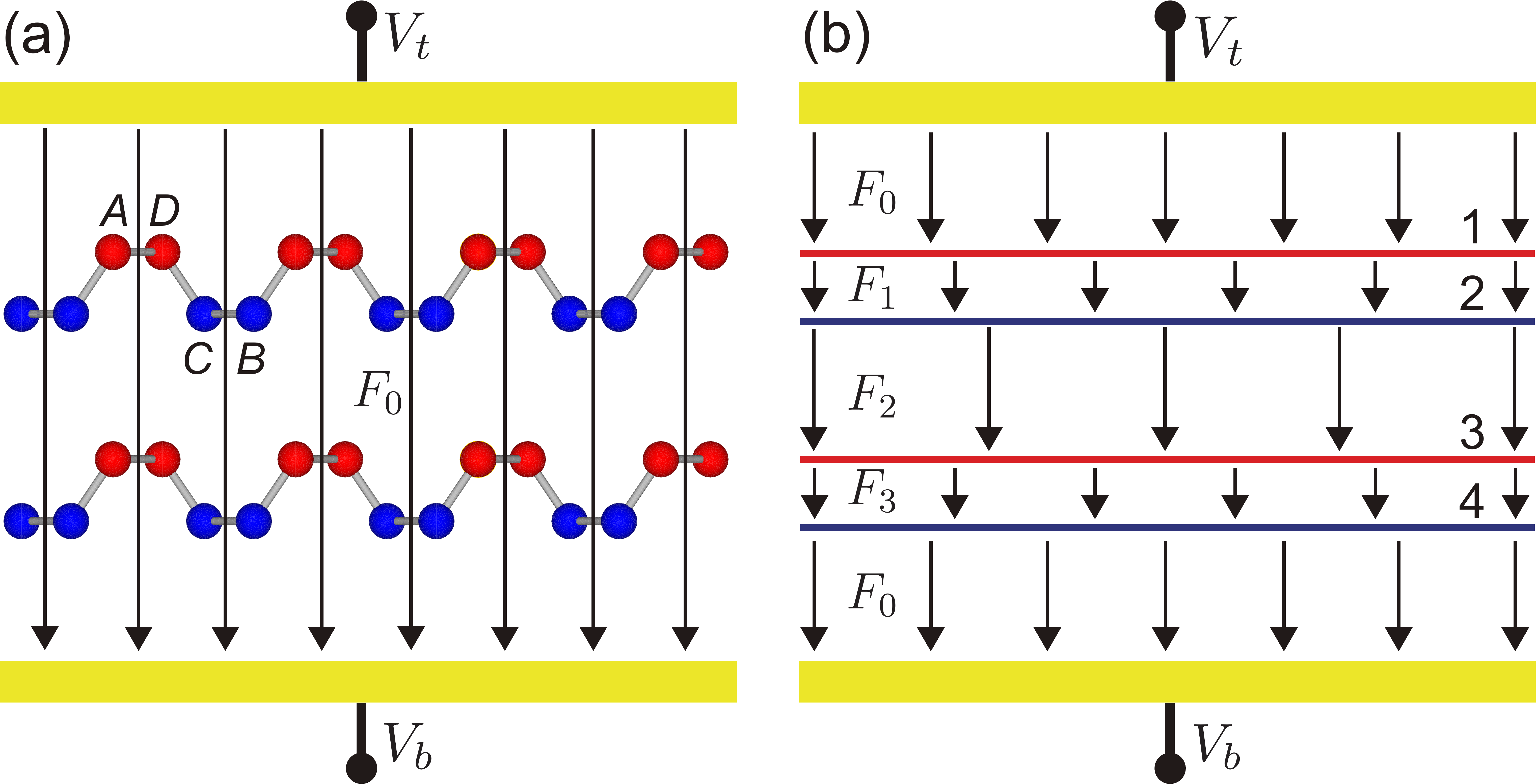}
\caption{(a) and (b) Sketch of bilayer phosphorene in the presence of a perpendicular electric field $F_0$, which can be produced by applying external (top and bottom) gate voltages ($V_t$ and $V_b$). $A$, $B$, $C$, and $D$ represent the four different sublattices in the unit cell of phosphorene. Because phosphorene has a puckered lattice structure, there are four atomic sublayers in bilayer phosphorene. Due to the field-induced charge screening, the applied electric field $F_0$ is no longer uniform across the structure, which turns into the screened electric fields $F_j$ ($j=1, 2, 3$) formed between the four atomic sublayers.
}\label{fig1}
\end{figure}

We consider undoped $AB$-stacked few-layer phosphorene in the presence of a perpendicular electric field $F_0$, as shown in Fig. \ref{fig1}. The perpendiducular electric field can be produced by applying external gate voltages as realized in recent experiments \cite{pengMidinfraredElectroopticModulation2017c, sherrottElectricalControlLinear2017a}. From density-functional theory (DFT) calculations \cite{cakirSignificantEffectStacking2015b,rudenkoRealisticDescriptionMultilayer2015a}, we know that this type of layer stacking is energetically the most stable for few-layer phosphorene.

The TB Hamiltonian for \textit{pristine} few-layer phosphorene is given by \cite{rudenkoRealisticDescriptionMultilayer2015a}
\begin{equation}\label{e1}
H=\sum_{i}\varepsilon_ic_i^{\dag}c_i+\sum_{i\neq j}t_{ij}^{\|}c_i^{\dag}c_j+\sum_{i\neq j}t_{ij}^{\perp}c_i^{\dag}c_j,
\end{equation}
where the summation runs over all lattice sites, $\varepsilon_i$ is the on-site energy at site $i$, $t_{ij}^{\|}$ ($t_{ij}^{\perp}$) is the intralayer (interlayer) hopping energy between sites $i$ and $j$, and $c_i^{\dag}$ ($c_j$) is the creation (annihilation) operator of an electron at site $i$ ($j$). For simplicity, the on-site energy $\varepsilon_i$ is set to zero for all the lattice sites. The Hamiltonian (\ref{e1}) is constructed on the basis of the $p_z$ orbital of the phosphorus atom. It was shown in Ref. \cite{rudenkoRealisticDescriptionMultilayer2015a} that (i) the $p_z$ orbital has a predominant contribution to the band structure of few-layer phosphorene in the low-energy region and (ii) with ten intralayer and five interlayer hopping parameters, this TB Hamiltonian can reproduce the band structure of few-layer phosphorene in the low-energy region as obtained by DFT-GW calculations. The ten intralayer hopping parameters (in units of eV) are $t_{1}^{\|}=-1.486$, $t_{2}^{\|}=+3.729$, $t_{3}^{\|}=-0.252$, $t_{4}^{\|}=-0.071$, $t_{5}^{\|}=+0.019$, $t_{6}^{\|}=+0.186$, $t_{7}^{\|}=-0.063$, $t_{8}^{\|}=+0.101$, $t_{9}^{\|}=-0.042$, $t_{10}^{\|}=+0.073$, and the five interlayer hopping parameters (in units of eV) are $t_{1}^{\perp}=+0.524$, $t_{2}^{\perp}=+0.180$, $t_{3}^{\perp}=-0.123$, $t_{4}^{\perp}=-0.168$, $t_{5}^{\perp}=+0.005$ \cite{rudenkoRealisticDescriptionMultilayer2015a}. These hopping parameters were illustrated in Figs. \ref{fig1}(c) and \ref{fig1}(d) of Ref. \cite{liTuningElectronicProperties2018c}.

The effect of a \textit{perpendicular electric field} is included into the TB Hamiltonian (\ref{e1}) by introducing an electrostatic potential to the on-site term, i.e., $\varepsilon_i \to \varepsilon_i + U_i$, with $U_i$ the electrostatic potential energy at site $i$, which is sublayer-dependent in few-layer phosphorene (i.e., $U_i=U_{l}$ for all sites $i$ in the sublayer $l$). It is known that applying an external electric field perpendicular to a few-layer system induces a charge redistribution over the stacking layers, which produces an internal electric field that counteracts the externally applied one (i.e., the electric-field-induced charge screening). As a consequence, the screened electric field across the few-layer structure is no longer uniform as the externally applied one and the magnitude of the screened electric field depends on the charge densities on the individual stacking layers. In order to take account of the field-induced charge screening in few-layer phosphorene, a self-consistent tight-binding approach can be employed and the electrostatic potential energy can be obtained within a self-consistent Hartree approximation. The details of how to obtain the self-consistent electrostatic potential energy were given in our previous work \cite{liTuningElectronicProperties2018c}. In the absence of the field-induced charge screening, the electrostatic potential energy can be modeled as a linear potential energy on the sublayer, i.e., $U_i=U_{l}=eF_0z_{l}$, with $z_{l}$ the sublayer position along the $z$ direction.

By diagonalizing the TB Hamiltonian (\ref{e1}) without (with) considering the field-induced charge screening in momentum space, we obtain the energy spectrum and the wave function of the unscreened (screened) electronic states in few-layer phosphorene subjected to a perpendicular electric field. Once the energy spectrum and the wave function are obtained, the optical conductivity tensor of few-layer phosphorene can be readily calculated by using the standard Kubo formula \cite{lowTunableOpticalProperties2014c}, which is given by
\begin{equation*}\nonumber
\sigma_{\alpha\beta}(\omega) = \frac{g_se^2\hbar}{4\pi^2i}\sum_{m, n}\int d\textbf{k} \frac{f[E_m(\textbf{k})] - f[E_n(\textbf{k})]}{E_m(\textbf{k}) - E_n(\textbf{k})}
\end{equation*}
\begin{equation}\label{e2}
\times \frac{\bra{\Psi_m(\textbf{k})}{\textbf{v}_{\alpha}}\ket{\Psi_n(\textbf{k})}
\bra{\Psi_n(\textbf{k})}{\textbf{v}_{\beta}}\ket{\Psi_m(\textbf{k})}}{E_m(\textbf{k}) - E_n(\textbf{k}) + \hbar\omega + i\eta},
\end{equation}
where $\alpha, \beta = x, y$ are the tensor indices, $\omega$ is the optical frequency, $g_s=2$ is the spin-degeneracy factor, $m, n$ are the band indices, $\textbf{k}=(k_x, k_y$) is the 2D wave vector, $f(E)$ is the Fermi-Dirac function, $E_{m,n}(\textbf{k})$ are the band energies, $\Psi_{m,n}(\textbf{k})$ are the wave functions, $\textbf{v}_{\alpha, \beta}$ are the $\alpha$ and $\beta$ components of the velocity operator $\textbf{v}=\hbar^{-1}\partial H/\partial \textbf{k}$, and $\eta$ is a finite broadening. There are four components in the optical conductivity tensor (\ref{e2}): $\sigma_{xx}$ and $\sigma_{yy}$ for the longitudinal optical conductivities and $\sigma_{xy}$ and $\sigma_{yx}$ for the transverse optical conductivities (i.e., the optical Hall conductivities). For the longitudinal components we have $\sigma_{xx}=\sigma_{yy}$ ($\sigma_{xx}\neq\sigma_{yy}$) for an isotropic (anisotropic) system. When calculating the optical conductivity tensor Eq. (\ref{e2}), we take the broadening $\eta=10$ meV and the temperature $T=300$ K. The real parts of the longitudinal and transverse optical conductivities are related to the optical absorption effect and optical Hall effect (e.g., Faraday rotation), respectively.

\begin{figure}
\center
\includegraphics[width=0.49\textwidth]{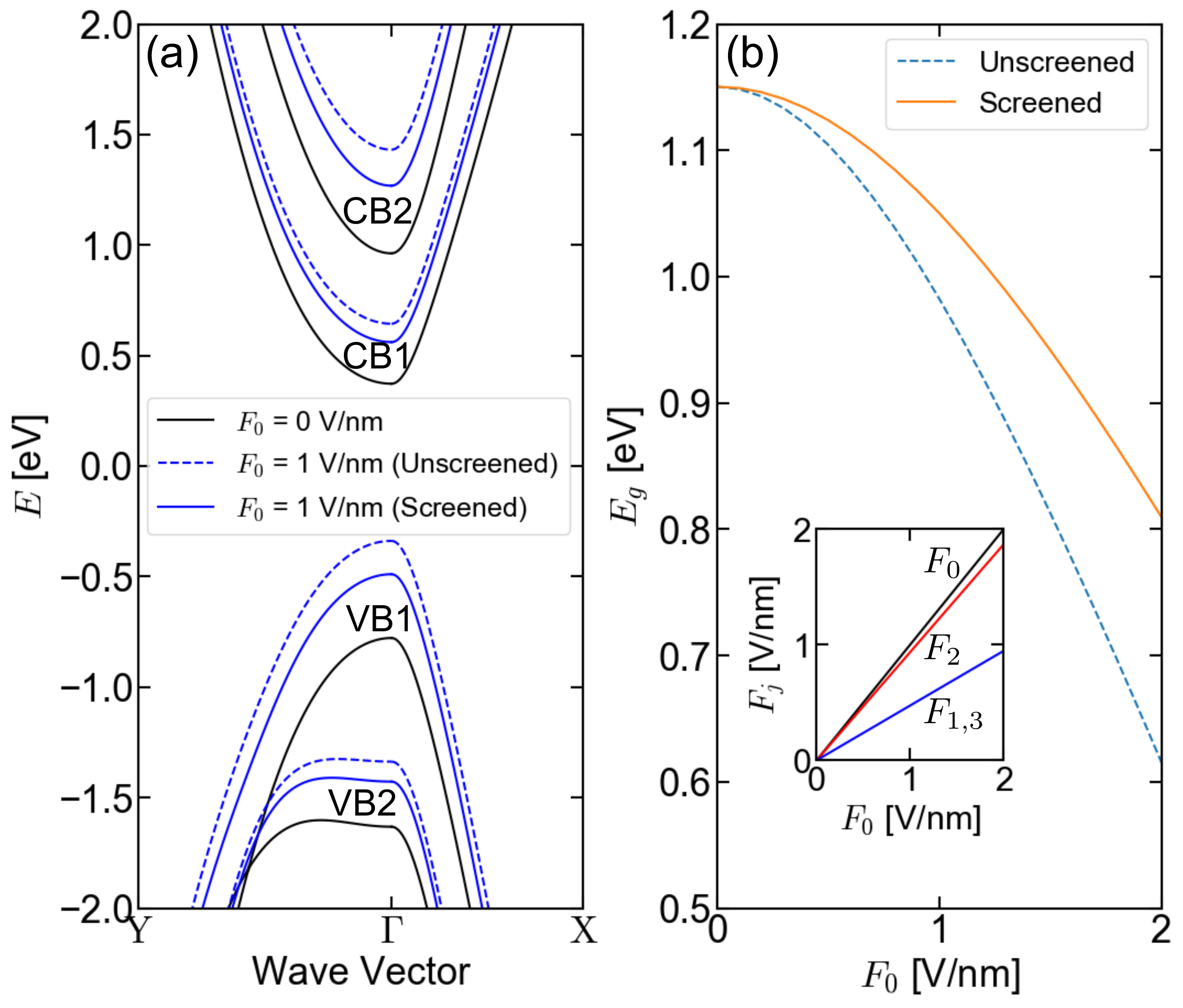}
\caption{Band structure (a) and band gap (b) of bilayer phosphorene in the absence/presence of a perpendicular electric field and without/with the field-induced charge screening. The labels CB1 and CB2 (VB1 and VB2) denote the $1^{th}$ and $2^{nd}$ conduction (valence) bands, respectively. The inset shown in (b) plots the screened electric fields as a function of the applied electric field. Here $F_0$ is the strength of the applied electric field and $F_1$, $F_2$, $F_3$ are the strengths of the screened electric fields [see Fig. \ref{fig1}(b)].}\label{fig2}
\end{figure}

\section{Band Structure}

The self-consistent TB calculations of the band structure of few-layer phosphorene in the presence of a perpendicular electric field are performed by taking account of the field-induced charge screening, which was shown to be of significant importance in biased few-layer phosphorene \cite{liTuningElectronicProperties2018c}.

In Fig. \ref{fig2}, we show the band structure and the band gap of bilayer phosphorene in the absence/presence of a perpendicular electric field and without/with the field-induced charge screening: (a) for the band structure and (b) for the band gap. In Fig. \ref{fig2}(a), $\Gamma=(0, 0)$ is the center of the Brillouin zone (BZ), and $\textrm{X}=(\pi/a, 0)$ and $\textrm{Y}=(0, \pi/b)$ are the two high-symmetry points in the BZ, with $a=4.43$ {\AA} and $b=3.37$ {\AA} being the side lengths of the rectangle unit cell of phosphorene. As can be seen, in the absence of the field-induced charge screening, the bilayer band structure is significantly changed by the applied electric field: the conduction bands (CBs) and the valence bands (VBs) are significantly shifted in energy due to the stark effect. Here the labels CB1 and CB2 (VB1 and VB2) in the band structure indicate the $1^{st}$ and $2^{nd}$ CBs (VBs), respectively. The field-induced change of the band structure in bilayer phosphorene is mainly due to the interlayer interaction and its competition with the electrostatic potential. Additionally, for bilayer phosphorene there is a crossing between the two VBs in the zero-field case, which is removed by the electric field. This is because the electric field breaks the inversion (layer) symmetry of bilayer phosphorene, which results in the interaction of the two VBs and thus the removal of their crossing. When the field-induced charge screening is taken into account, the bilayer band structure is further changed, as shown in Fig. \ref{fig2}(a): the CBs and the VBs are further shifted in energy due to the screening effect. We also find that the electric-field tuning of the band gap of bilayer phosphorene is distinctively different in the presence and absence of the field-induced charge screening. As can be seen in Fig. \ref{fig2}(b), the unscreened band gap of bilayer phosphorene decreases dramatically with increasing electric-field strength, wheareas the magnitude of this band-gap decrease is significantly reduced in the presence of the field-induced charge screening. This is not surprising because the screened electric fields ($F_1$, $F_2$, and $F_3$) have a smaller strength than the externally applied one ($F_0$), as shown in the inset of Fig. \ref{fig2}(b).

\section{Linear Dichroism}

Because we consider undoped few-layer phosphorene in the presence of a perpendicular electric field, there are no excess charge carriers (electrons and holes) in the system and so the Fermi energy of the system is located within the band-gap region of the system \cite{liTuningElectronicProperties2018c}. In this situation, when the system is excited by polarized light, only interband transitions from the occupied valence bands to the unoccupied conduction bands contribute to the optical conductivity of the system. For this reason, we only calculate the interband optical conductivity of undoped few-layer phosphorene in the presence of a perpendicular electric field. In order to investigate the linear dichroism of few-layer phosphorene, the interband optical conductivities for $x$ and $y$ polarizations (the longitudinal components) are calculated.

\begin{figure}
\center
\includegraphics[width=0.49\textwidth]{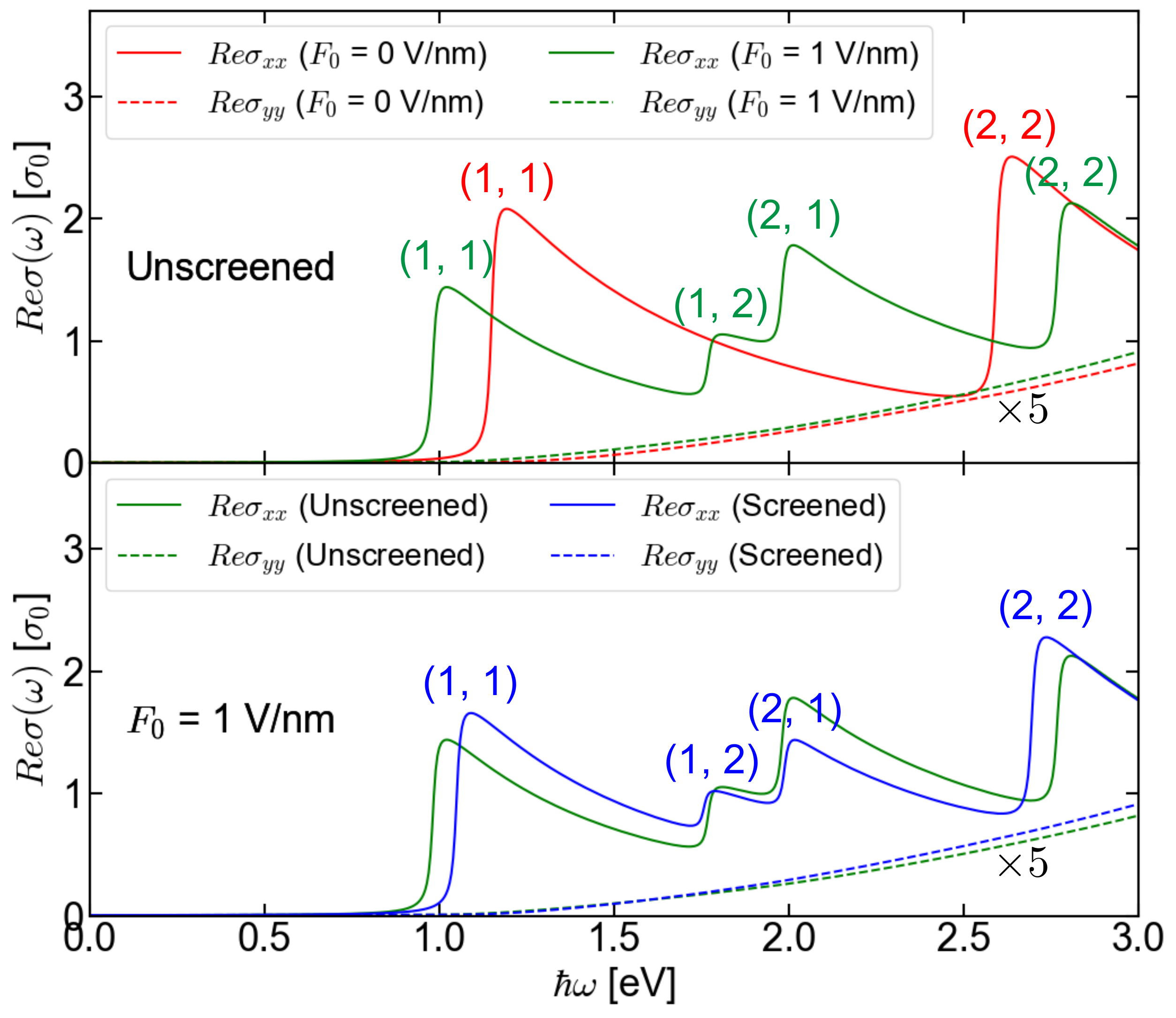}
\caption{Real parts of the interband optical conductivities $Re\sigma_{xx}$ and $Re\sigma_{yy}$ of bilayer phosphorene in the absence/presence of a perpendicular electric field and without/with the field-induced charge screening. The values of $Re\sigma_{yy}$ are enlarged with a factor of 5 to make the curves more visible. Here $\sigma_0=e^2/(4\hbar)$ is the universal optical conductivity of graphene and $(m, n)$ indicates the interband transition from the $m^{th}$ VB to the $n^{th}$ CB.}\label{fig3}
\end{figure}

We also calculate the band structures and the band gaps of monolayer and trilayer phosphorene in the presence of a perpendicular electric field. When compared to the band structure and the band gap of bilayer phosphorene, we find that those of monolayer (trilayer) phosphorene are less (more) affected by the applied electric field and the field-induced charge screening, such as the smaller (larger) stark shift of the band gap. This is because there are both intralayer and interlayer electronic interactions and charge screenings in multilayer phosphorene whereas only intralayer ones are present in monolayer phosphorene \cite{liTuningElectronicProperties2018c}. Our numerical calculations also indicated that with increasing number of stacking layers, the applied electric field and the field-induced charge screening have a more pronounced effect on the band structure and the band gap \cite{liTuningElectronicProperties2018c}.

\begin{figure*}[htbp!]
\center
\includegraphics[width=0.99\textwidth]{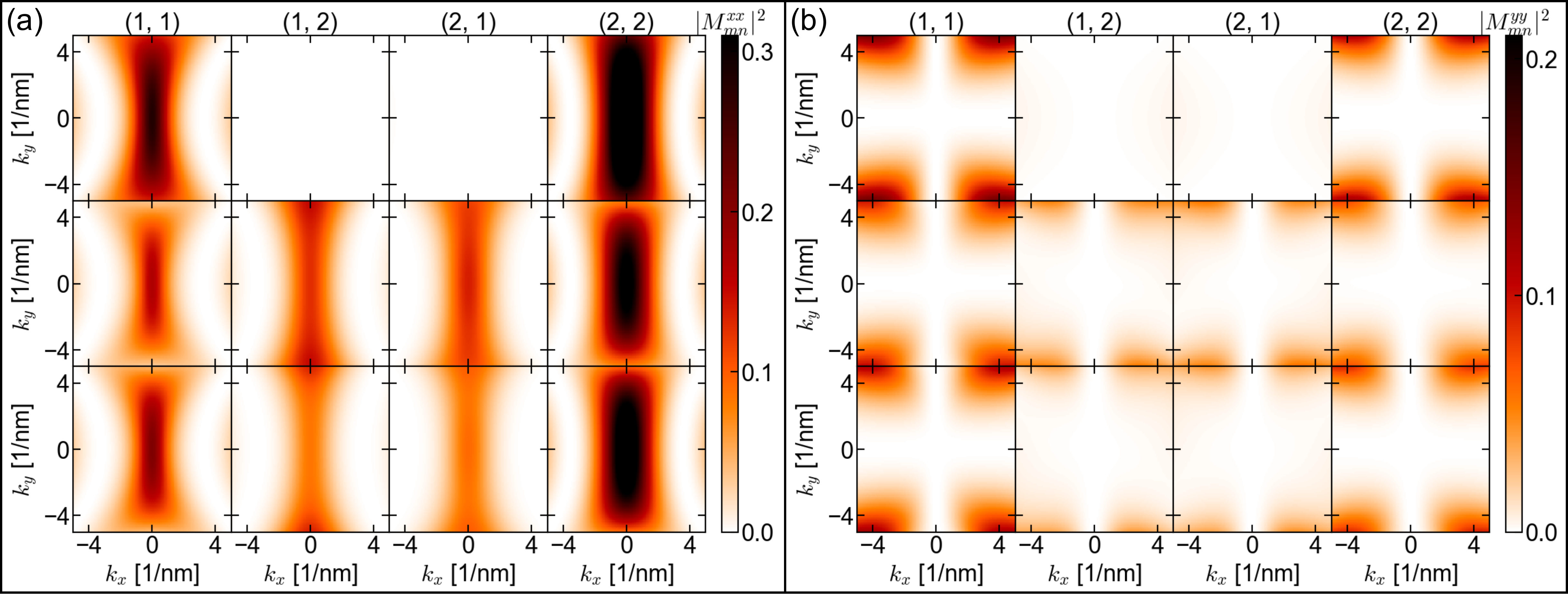}
\caption{Interband momentum matrix elements $|M_{mn}^{xx}|^2$ (a) and $|M_{mn}^{yy}|^2$ (b) in the absence/presence of a perpendicular electric field and without/with the field-induced charge screening. The top, middle, and bottom panels in both (a) and (b) show the results for the cases of $F_0=0$ V/nm, $F_0=1$ V/nm without screening, and $F_0=1$ V/nm with screening, respectively.}\label{fig4}
\end{figure*}

In Fig. \ref{fig3}, we show the real-part interband optical conductivities $Re\sigma_{xx}$ and $Re\sigma_{yy}$ of bilayer phosphorene in the absence/presence of a perpendicular electric field and without/with the field-induced charge screening. Here, we use the index pair $(m, n)$ to denote the interband transition from the $m^{th}$ VB to the $n^{th}$ CB. As can be seen in Fig. \ref{fig3}, distinct absorption peaks appear in $Re\sigma_{xx}$ due to the strong quantum confinement along the out-of-plane direction (i.e., along the $z$ direction). This confinement leads to the formation of quantized subbands along the $z$ direction. The number of discrete subbands ($N_S$) depends on the number of stacking layers ($N_L$) as $N_S=4N_L$. In the absence of the electric field, there are only two absorption peaks in $Re\sigma_{xx}$, which are associated with the interband transitions $(1,1)$ and $(2,2)$; whereas in the presence of the electric field, two more absorption peaks are induced in $Re\sigma_{xx}$, which correspond to the interband transitions $(1, 2)$ and $(2, 1)$. However, when the field-induced charge screening is taken into account, the magnitudes of these two extra absorption peaks are reduced. In addition, the cut-off absorption of $Re\sigma_{xx}$ is shifted to the lower photon energy (i.e., red-shifted) by the electric field.

In order to understand the appearance of the two extra absorption peaks in $Re\sigma_{xx}$ induced by the perpendicular electric field and the reduction of their magnitudes induced by the field-induced charge screening, we need to look into the interband momentum matrix element for the $x$ polarization, defined as
$|M_{mn}^{xx}(\textbf{k})|^2=|\bra{\Psi_m(\textbf{k})}{\partial H}/{\partial k_x} \ket{\Psi_n(\textbf{k})}|^2$,
in the absence/presence of a perpendicular electric field and without/with the field-induced charge screening. The results of $|M_{mn}^{xx}|^2$ for the interband transitions $(1, 1)$, $(2, 2)$, $(1, 2)$ and $(2, 1)$ are shown in Fig. \ref{fig4}(a) for the absence/presence of a perpendicular electric field and without/with the field-induced charge screening. It is clear that $|M_{12}^{xx}|^2$ and $|M_{21}^{xx}|^2$ are vanishing in the absence of the electric field whereas they become finite in the presence of the electric field, and their magnitudes are reduced by the field-induced screening. Therefore, we can establish the following optical selection rules in bilayer phosphorene: (i) The interband transitions $(1, 1)$ and $(2, 2)$ are optically allowed in both the absence and presence of a perpendicular electric field; (ii) The interband transitions $(1, 2)$ and $(2, 1)$ are optically forbidden in the absence of a perpendicular electric field but they become optically allowed in the presence of a perpendicular electric field. Fundamentally, these optical selection rules are closely related to the parities (symmetries) of the electron (hole) wave functions in the conduction (valence) bands. As is known, the applied electric field breaks the inversion symmetry of bilayer phosphorene with respect to the $z=0$ plane, and consequently, the parities of the electron and hole wave functions along the $z$ direction are changed, which results in different overlaps of the electron and hole wave functions and so different magnitudes of the momentum matrix elements between them. One can understand this field-induced change of the wave-function parity (or symmetry) in a phosphorene bilayer in a similar way as that in a quantum well with an external electric field applied along the growth direction, where the electron and hole wave functions change from the Cosine/Sine-like functions with even/odd parities to the Airy-like functions with no parities \cite{mcilroyEffectElectricField1986}.

\begin{figure}
\center
\includegraphics[width=0.49\textwidth]{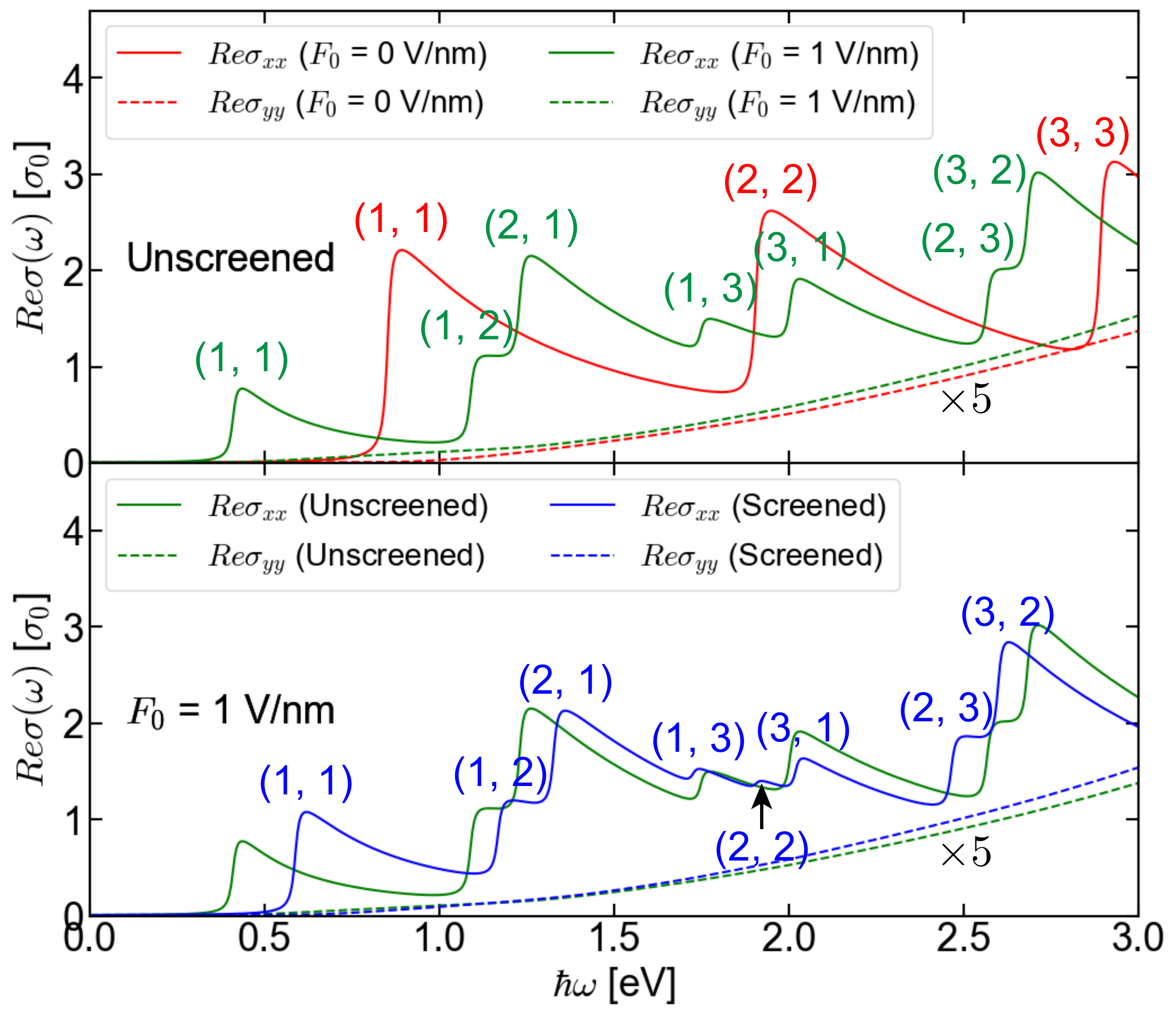}
\caption{Real parts of the interband optical conductivities $Re\sigma_{xx}$ and $Re\sigma_{yy}$ of trilayer phosphorene in the absence/presence of a perpendicular electric field and without/with the field-induced charge screening. The values of $Re\sigma_{yy}$ are enlarged by a factor of 5 to make the curves more visible.}\label{fig5}
\end{figure}

\begin{figure*}
\center
\includegraphics[width=0.93\textwidth]{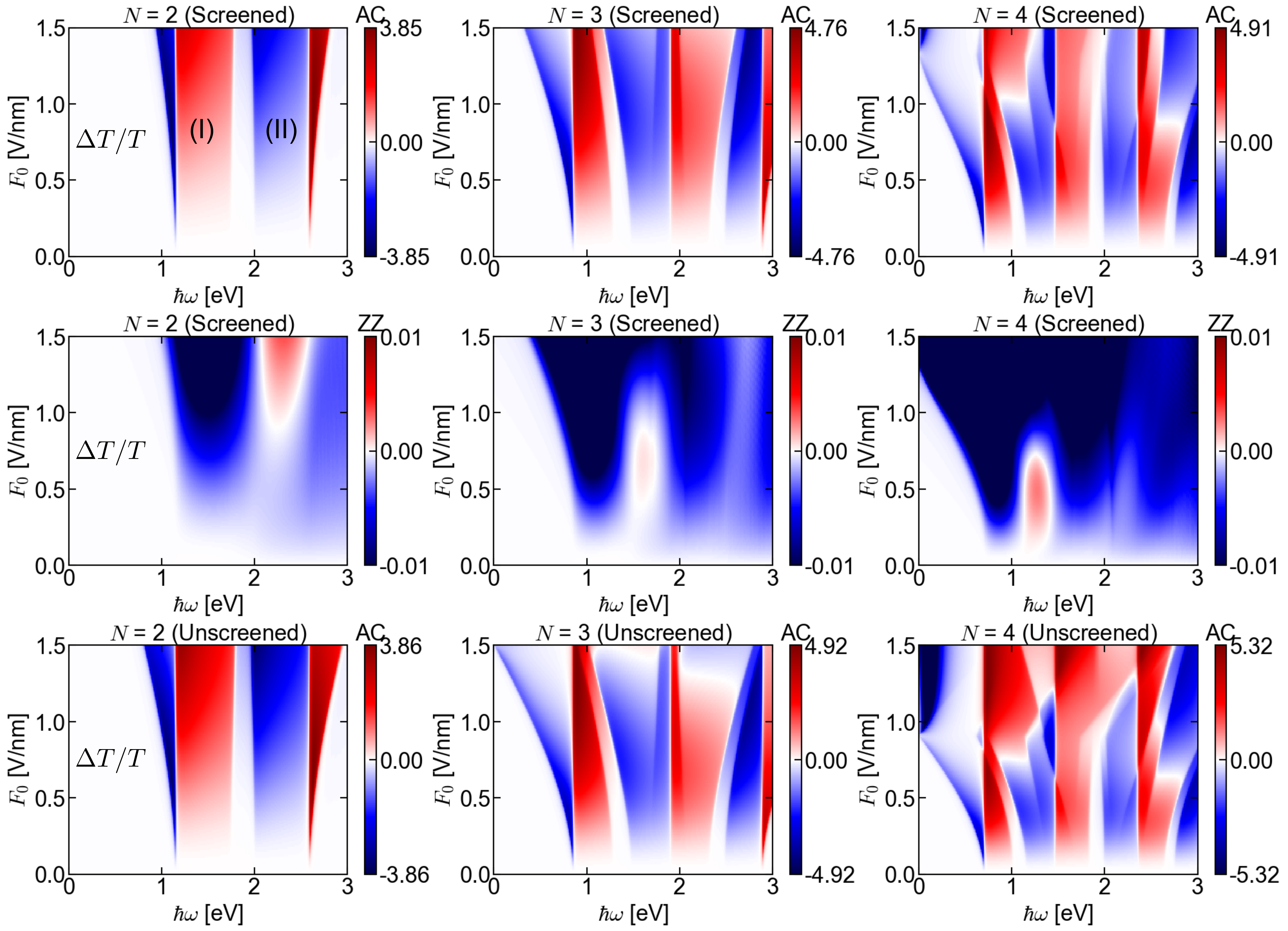}
\caption{Electric-field modulation $\Delta T/T$ of the optical transmission of $N$-layer phosphorene ($N=2, 3, 4$) for the AC (armchair) and ZZ (zigzag) polarizations. Here the values of $\Delta T/T$ as indicated by the colorbar are given in percentage .
}\label{fig6}
\end{figure*}

More interestingly, we observe in Fig. \ref{fig3} that there is a significant difference between $Re\sigma_{xx}$ and $Re\sigma_{yy}$, i.e., the real part of the interband optical conductivity exhibits a significant linear dichroism. As mentioned above, $Re\sigma_{xx}$ has distinct absorption peaks, which is strongly affected by the perpendicular electric field and is greatly influenced by the field-induced charge screening. However, there are no distinct absorption peaks in $Re\sigma_{yy}$ and so it is featureless. Moreover, it is almost unaffected by the applied electric field and the field-induced charge screening. In order to understand the absorption features of $Re\sigma_{yy}$, we need to look into the interband momentum matrix element for the $y$ polarization, defined as $|M_{mn}^{yy}(\textbf{k})|^2=|\bra{\Psi_m(\textbf{k})}{\partial H}/{\partial k_y} \ket{\Psi_n(\textbf{k})}|^2$, in the absence/presence of a perpendicular electric field and without/with the field-induced charge screening. The results of $|M_{mn}^{yy}|^2$ for the interband transitions $(1, 1)$, $(2, 2)$, $(1, 2)$ and $(2, 1)$ are shown in Fig. \ref{fig4}(b). As can be seen by comparing Fig. \ref{fig4}(a) and Fig. \ref{fig4}(b), $|M_{11}^{yy}|^2$ and $|M_{22}^{yy}|^2$ are significantly smaller than $|M_{11}^{xx}|^2$ and $|M_{22}^{xx}|^2$, especially around the $\Gamma$ point $(k_x, k_y)=(0, 0)$. Such strong anisotropy arises from the unique lattice structure of phosphorene, where phosphorus atoms form a puckered honeycomb lattice with very different electronic behaviours along the $x$ (armchair) and $y$ (zigzag) directions. Moreover, $|M_{11}^{yy}|^2$ and $|M_{22}^{yy}|^2$ as well as $|M_{12}^{yy}|^2$ and $|M_{21}^{yy}|^2$ are almost unaffected by the applied electric field and the field-induced charge screening. These results reflect well the highly optical anisotropy of bilayer phosphorene.

In Fig. \ref{fig5}, we show the real parts of the interband optical conductivities $Re\sigma_{xx}$ and $Re\sigma_{yy}$ of trilayer phosphorene in the absence/presence of a perpendicular electric field and without/with the field-induced screening. As can be seen, there are more absorption peaks in the conductivity spectrum of trilayer phosphorene when compared to that of bilayer phosphorene shown in Fig. \ref{fig3}. This is because the confinement energy is smaller and there are more discrete subbands in both the conduction and valence bands due to the increased number of stacking layers. Here we identify in the conductivity spectrum of trilayer phosphorene the interband transition channels $(1, 1)$, $(2, 2)$, $(3, 3)$ in the absence of the electric field and the field-induced channels $(1, 2)$, $(2, 1)$, $(1, 3)$, $(3, 1)$, $(2, 3)$, $(3, 2)$ in the presence of the electric field. The absence/presence of these interband transition channels, together with the decrease/increase of their magnitudes, can also be understood in terms of the interband momentum matrix elements of trilayer phosphorene, as done previously for bilayer phosphorene.

\section{Electric-field Modulation}

The electric-field modulation of the optical properties of few-layer phosphorene is investigated. This electro-optic modulation is broadly used to measure the change in the optical coefficients (such as the optical absorption, reflection and transmission) of a material sample due to the application of an external electric field. For thin-layer samples fabricated from e.g. 2D materials, a higher precision of such modulation is obtained by using optical transmission experiments. The reflection and transmission coefficients of polarized light through a thin-layer sample can be obtained by solving the Maxwell equations with appropriate boundary conditions for electromagnetic fields \cite{stauberOpticalConductivityGraphene2008}. For the case of normal incidence, the optical transmission coefficient is given by \cite{stauberOpticalConductivityGraphene2008}
\begin{equation}\label{e3}
T(\omega)=\sqrt{\frac{\epsilon_2}{\epsilon_1}}\frac{4(\epsilon_1\epsilon_0)^2}
{|(\sqrt{\epsilon_1\epsilon_2}+\epsilon_1)\epsilon_0+\sqrt{\epsilon_1}Re\sigma(\omega)/c|^2}
\end{equation}
where $\epsilon_0$ is the permittivity of vacuum, $\epsilon_1$ and $\epsilon_2$ are the relative permittivities of media on the top and bottom of a thin-layer sample, and $c$ is the speed of light in vacuum. In the present work, we take $\epsilon_1=\epsilon_2=1$ which models a free-standing sample. The electric-field modulation strength of the optical transmission is defined as $\Delta T/T=\big[T_F(\omega)-T_0(\omega)\big]/T_0(\omega)$, with the subscripts $F$ and $0$ referring to the presence and absence of an externally applied field, respectively.

In Fig. \ref{fig6}, we show the electric-field modulation strengths $\Delta T/T$ of the optical transmission of $N$-layer phosphorene ($N=2, 3, 4$) for the armchair ($x$) and zigzag ($y$) polarizations, denoted by AC and ZZ, respectively. Here the values of $\Delta T/T$ as indicated in the colorbar are given in percentage. As can be seen, $\Delta T/T$ is much larger for the AC polarization than for the ZZ polarization, which indicates that the field modulation strength for the ZZ polarization is negligibly small. For both the AC and ZZ polarizations, $\Delta T/T$ oscillates as a function of the photon energy, with signs changed between negative (blue color) and positive (red color). And both of them can be divided into different regions with negative and positive values. For illustration purposes, two distinct regions are labeled by (I) and (II) in the far-top-left panel of Fig. \ref{fig6}. By analysing the field dependence of the interband optical conductivity of bilayer phosphorene, we find that the four different regions are induced by different interband transitions. For instance, a positive value of $\Delta T/T$ in region (I) is mainly induced by the interband transition $(1, 1)$, while a negative value of $\Delta T/T$ in region (II) is mainly induced by the interband transition $(2, 1)$. This is because the momentum matrix element of the zero-field interband transition $(1, 1)$ [$(2, 1)$] can be reduced (enhanced) by applying a perpendicular field, as described before, and consequently the corresponding optical transmission can be enhanced (reduced) by the applied field, thereby leading to a positive (negative) value of the field modulation strength. Notice that the field modulation for the AC polarization becomes more efficient with increasing number of phosphorene layers, as seen by comparing $\Delta T/T$ for the cases of $N=2, 3, 4$. This is reflected by the fact that the modulation strength becomes larger and the modulation range becomes wider. For instance, the absolute value of $\Delta T/T$ increases from $3.85\%$ in the two-layer case to $4.91\%$ in the four-layer case, and the energy range of $\Delta T/T$ extends from 1$-$3 eV in the two-layer case to 0$-$3 eV in the four-layer case. Our numerical results indicate that with increasing field strength from 0 to 1.5 V/nm, four-layer phosphorene undergoes a semiconductor-to-semimetal transition, leading to a zero band gap in the band structure. Furthermore, we find that the field modulation strength in the presence of the field-induced charge screening is smaller when compared to that in the absence of the field-induced charge screening. This is because the strength of the applied field is reduced by the field-induced charge screening.

\begin{figure}
\center
\includegraphics[width=0.49\textwidth]{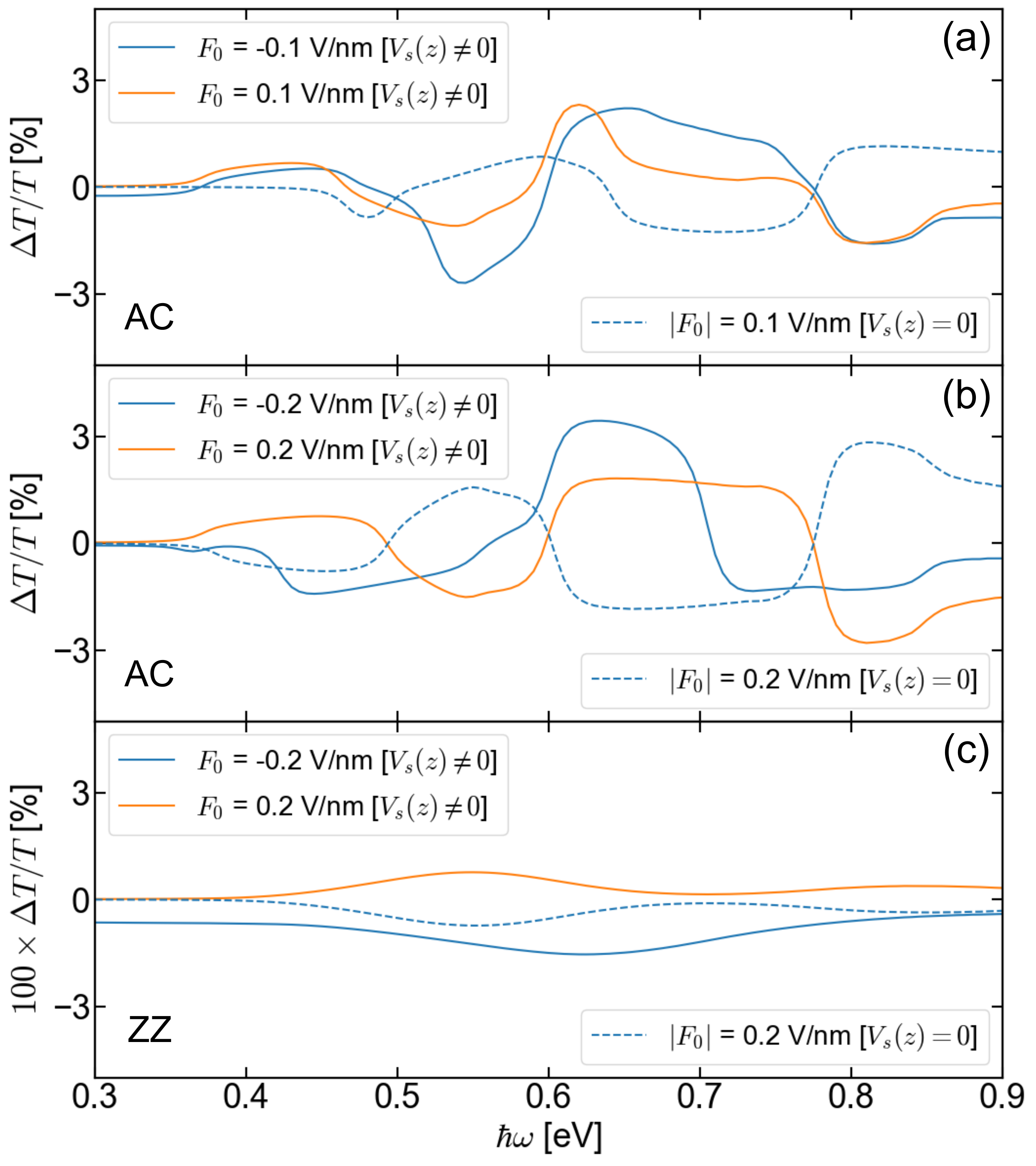}
\caption{Modulation strength $\Delta T/T$ of $p$-doped 7-layer phosphorene with a hole density of $5\times 10^{12}$ cm$^{-2}$ for opposite directional electric fields ($\pm F_0$): (a) $F_0=0.1$ V/nm in the absence/presence of $V_s(z)$ for the AC polarization; (b) $F_0=0.2$ V/nm in the absence/presence of $V_s(z)$ for the AC polarization; (c) $F_0=0.2$ V/nm in the absence/presence of $V_s(z)$ for the ZZ polarization. Here the results are shown in the presence of the field-induced charge screening and the values of $\Delta T/T$ for the ZZ polarization are enlarged by a factor of 100 for better visualization.}\label{fig7}
\end{figure}


Recent experiments reported the electric-field modulation of the linear dichroism of few-layer phosphorene \cite{pengMidinfraredElectroopticModulation2017c, sherrottElectricalControlLinear2017a}. In agreement with our theoretical results shown in Fig. \ref{fig6}, they found that (i) the field modulation oscillates significantly with positive and negative values as a function of the photon energy for the AC polarization, and (ii) the modulation strength is negligibly small (almost no modulation within experimental resolution) for the ZZ polarization. Moreover, as observed in the experiment \cite{pengMidinfraredElectroopticModulation2017c, sherrottElectricalControlLinear2017a}, the modulation strength for opposite electric-field directions exhibits an asymmetry with respect to the zero electric field (or zero gate voltage), which was explained by the presence of a small built-in electric field at zero gate voltage due to the substrate-induced hole doping in realistic few-layer phosphorene samples. Motivated by this, we calculate the modulation strength of a 7-layer phosphorene sample (corresponding to the layer thickness of 3.5 nm as determined in the experiment \cite{sherrottElectricalControlLinear2017a}) with an assumed hole density of $n_h=5\times 10^{12}$ cm$^{-2}$ for opposite electric-field directions. In order to simulate the substrate-induced potential, we also assume a substrate-induced charge distribution, which is given by $\rho(z)=n_s\delta(z)$ with $n_s=n_h$ due to the charge neutrality condition. By solving the Poisson equation $d^2V_s(z)/dz^2=e^2\rho(z)/(\epsilon_0\kappa)$, the substrate-induced potential $V_s(z)$ can be obtained. Here $\kappa$ is the dielectric constant of few-layer phosphorene which we take $\kappa=6$ for 7-layer phosphorene, which is close to the values obtained by DFT calculations \cite{DielectricConstantPhosphorene}.
The calculation results are shown in Fig. \ref{fig7}. As can be seen, the modulation strength indeed exhibits an asymmetry for opposite field directions in the presence of the substrate-induced potential (full curves), and the magnitude of this asymmetry becomes more pronounced with increasing field strength. However, in the absence of the substrate-induced potential (dashed curves), the modulation asymmetry between opposite field directions disappears, as intuitively expected. This is not surprising because the effective field strength becomes $F_0+F_1(z)$ [$F_0-F_1(z)$] for the one (other) field direction with $F_1(z)$ being the built-in field due to the substrate potential $V_s(z)$. Furthermore, the field modulation asymmetry is much larger for the AC polarization than for the ZZ polarization, which agrees with the experiment \cite{pengMidinfraredElectroopticModulation2017c,
sherrottElectricalControlLinear2017a}.

\section{Faraday Rotation}

When shining a beam of light on an optical medium in the presence of a magnetic field, the light polarization can be rotated in two different configurations, i.e., the Kerr rotation when reflected at the surface of the medium and the Faraday rotation when transmitted through the interior of the medium. These rotations are a direct consequence of the optical Hall effect arising from the carrier-photon interaction in the presence of an external magnetic field. The optical Hall effect (e.g. the Faraday rotation) has been explored in ordinary and graphene quantum Hall systems \cite{OpticalHallEffectinGraphene,OpticalHallEffectin2DEG}. Here we show the the Faraday rotation in few-layer phosphorene in the absence of an external magnetic field. This is motivated by the fact that few-layer phosphorene has a reduced lattice symmetry due to its puckered lattice structure.

\begin{figure}
\center
\includegraphics[width=0.49\textwidth]{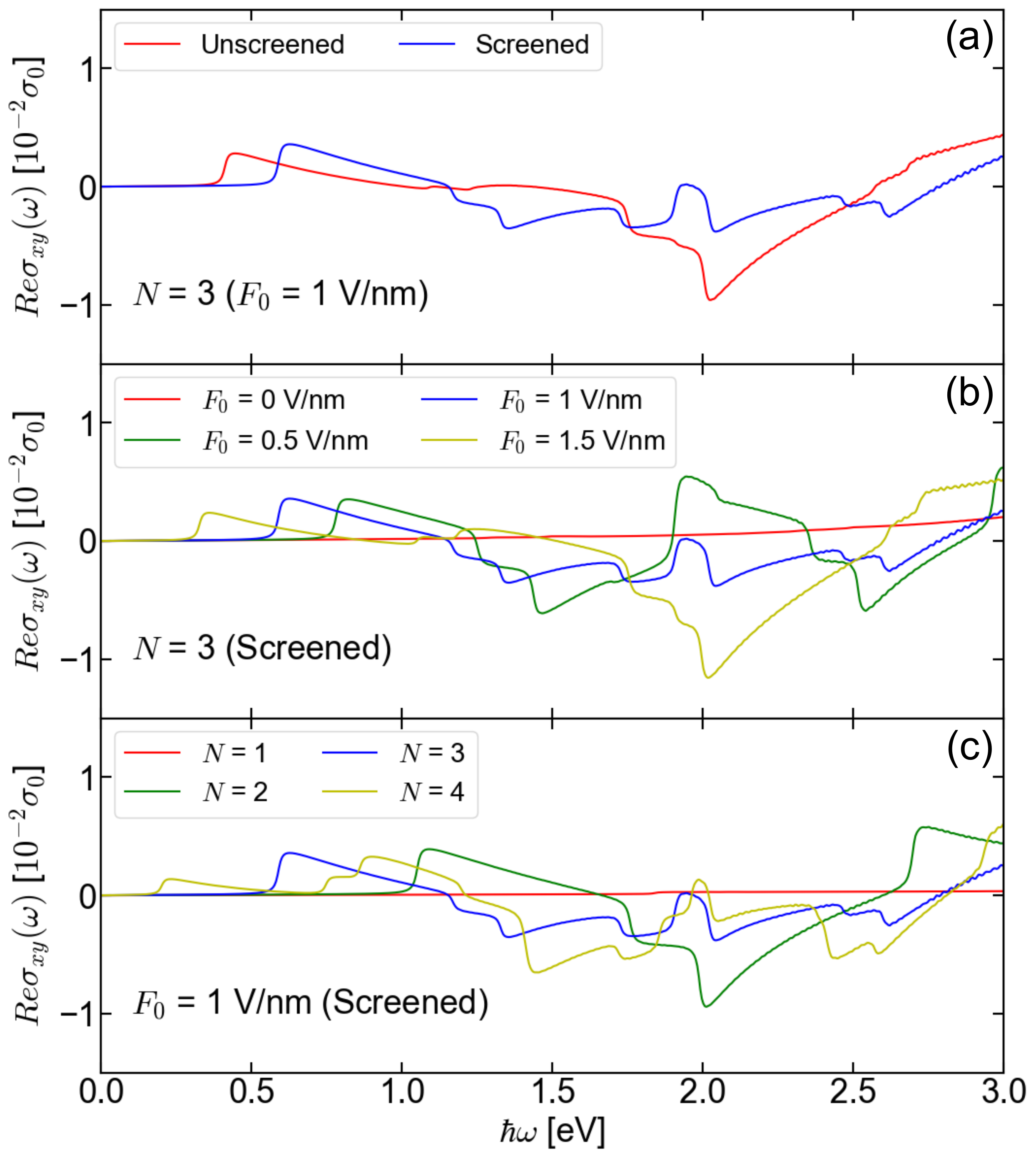}
\caption{Real parts of the interband optical Hall conductivities $Re\sigma_{xy}$ and $Re\sigma_{yx}$ of $N$-layer phosphorene: (a) Effect of the field-induced charge screening for the field strength of $F_0=1$ V/nm in trilayer phosphorene; (b) Dependence on the field strength $F_0$ with the screening effect in trilayer phosphorene; (c) Dependence on the layer number $N$ with the screening effect for the field strength of $F_0=1$ V/nm.}\label{fig8}
\end{figure}

In Fig. \ref{fig8}, we show the real-part optical Hall conductivity $Re\sigma_{xy}$ of few-layer phosphorene: (a) Effect of the field-induced charge screening for the field strength of $F_0=1$ V/nm in trilayer phosphorene; (b) Dependence on the field strength $F_0$ with the screeening effect in trilayer phosphorene; (c) Dependence on the number of layers with the screeening effect for the field strength of $F_0=1$ V/nm. As can be seen, $Re\sigma_{xy}$ is greatly influenced by the field-induced charge screening and is strongly dependent on the field strength and on the layer number. Interestingly, $Re\sigma_{xy}$ has either positive or negative values depending on the optical frequency. According to the relation $\theta_F\simeq Re\sigma_{xy}/[(1+n_{sub})c\epsilon_0]$ with $\theta_F$ the Faraday rotation angle and $n_{sub}$ the refractive index of the substrat \cite{OpticalHallEffectin2DEG}, the frequency dependence of $Re\sigma_{xy}$ indicates that the polarization of light can be rotated in opposite directions by varying the frequency of light when transmitted through few-layer phosphorene. Notice that the optical Hall conductivity is smaller than its longitudinal counterpart by two orders of magnitude (i.e., $100Re\sigma_{xy} \sim Re\sigma_{xx}$), which agrees with the results obtained from DFT calculations for other 2D materials such as GaS and GaSe multilayers \cite{liThicknessdependentMagnetoopticalEffects2018}. The results shown here, i.e., the strong dependence of the optical Hall conductivity on the perpendicular electric field and on the number of phosphorene layers, indicate that Faraday rotation is not only electrically tunable but also layer-thickness dependent.

\section{Discussions and Conclusions}

Before closing this paper, we present the following additional remarks:

(i) The results obtained for the optical properties of few-layer phosphorene are generic for an arbitrary number of phosphorene layers. The main differences when increasing the number of phosphorene layers are that the cut-off absorption is red-shifted due to the reduced band gap, the number of absorption peaks increases due to the appearance of more discrete subbands, and that the electric-field modulation of optical absorption becomes more pronounced.

(ii) The optical conductivity (\ref{e2}) is calculated on the basis of single-particle states, implying that excitonic effects are neglected. Such many-body effects will result in additional excitonic absorption peaks below the band gap. However, the theoretical band gaps obtained by the single-particle TB approch were shown to agree with those obtained by the many-body DFT-GW method \cite{rudenkoRealisticDescriptionMultilayer2015a}. Moreover, recent experimental and theoretical studies of the infrared absorption spectrum of few-layer phosphorene also indicated the validity of the single-particle approximation by comparing the experimental and theoretical results of the optical absorption spectrum \cite{zhangInfraredFingerprintsFewlayer2017b}.

(iii) The present work considered both undoped and doped few-layer phosphorene. It was shown \cite{liTuningElectronicProperties2018c} that when the perpendicular electric field is produced by only a top (bottom) gate, few-layer phosphorene will become $n$-doped ($p$-doped) with a finite density of electrons (holes) and as a consequence, the Fermi energy of the system will fall into the conduction (valence) band instead of being located within the band gap. In this situation, both interband and intraband transitions will contribute to the optical conductivity of few-layer phosphorene. The intraband optical conductivity is expected to be different for the $n$-doped and $p$-doped systems due to the electron-hole asymmetry of few-layer phosphorene. Therefore, the linear dichroism and the Faraday rotation induced by intraband transitions are expected to be different in the $n$-doped and $p$-doped few-layer phosphorene systems.

\begin{figure}
\center
\includegraphics[width=0.49\textwidth]{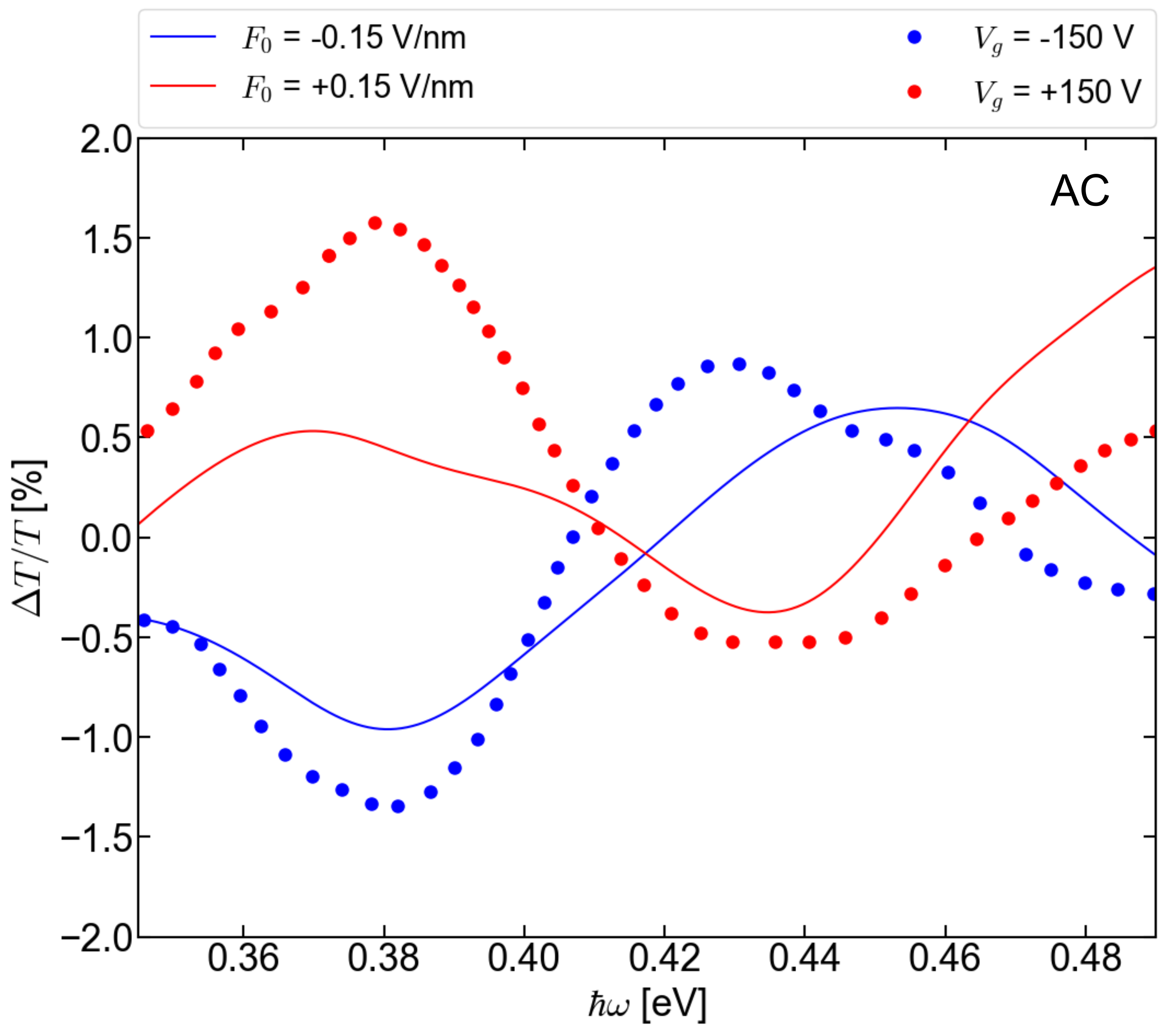}
\caption{Theoretical (curves) and experimental (dots) modulation strengths $\Delta T/T$ of a 17-layer phosphorene sample with a substrate-induced hole density of $n_h\sim10^{13}$ cm$^{-2}$ for different electric fields $F_0=\pm0.15$ V/nm (corresponding to the experimental gate voltages of $V_g=\pm150$ V). The results are shown for the AC polarization and the experiment data are taken from \cite{pengMidinfraredElectroopticModulation2017c}.}\label{fig9}
\end{figure}

(iv) We present now a direct comparison between theory and the experiment of Ref. \cite{pengMidinfraredElectroopticModulation2017c}. The device structure fabricated in \cite{pengMidinfraredElectroopticModulation2017c} consists of the active layer (few-layer BP), the encapsulation layer (Al$_2$O$_3$), and the substrate layer (SiO$_2$). We denote $d_{BP}$, $d_{Al_2O_3}$, and $d_{SiO_2}$ being the thicknesses of the BP layer, the Al$_2$O$_3$ layer, and the SiO$_2$ layer, respectively. In the experiment \cite{pengMidinfraredElectroopticModulation2017c} the thicknesses of the two dielectric layers were set to $d_{Al_2O_3}=10$ nm and $d_{SiO_2}=450$ nm. For a given gate voltage $V_g$ applied to the device structure, the corresponding electric field applied to the BP layer can approximately be computed as $
F_0=V_g\epsilon_{Al_2O_3}\epsilon_{SiO_2}/(\epsilon_{Al_2O_3}\epsilon_{SiO_2}d_{BP}
+\epsilon_{BP}\epsilon_{SiO_2}d_{Al_2O_3}+\epsilon_{BP}\epsilon_{Al_2O_3}d_{SiO_2})$,
with $\epsilon_{Al_2O_3}=9.5$, $\epsilon_{BP}=8.3$, and $\epsilon_{SiO_2}=3.9$ being the dielectric constants of the Al$_2$O$_3$ layer, the BP layer, and the SiO$_2$ layer, respectively. We consider a 17-layer BP sample in the presence of opposite electric fields $F_0=\pm0.15$ V/nm, corresponding to the layer thickness of $d_{BP}=9$ nm and the gate voltages of $V_g=\pm150$ V in the experiment \cite{pengMidinfraredElectroopticModulation2017c}. We assume a substrate-induced hole doping with a finite density of $n_h\sim 10^{13}$ cm$^{-2}$. Since the BP sample is $p$-doped due to the substrate-induced hole density, the Fermi energy is located within the valence subbands and thus both the interband and intraband transitions will have to be included in the calculation of the optical conductivity [Eq. (\ref{e2})]. Moreover, in the calculation the broadening factor appearing in Eq. (\ref{e2}) is taken as $\eta=20$ meV. We show in Fig. \ref{fig9} the direct comparison between theory and experiment for the AC polarization. As can be seen, the modulation pattern agrees qualitatively with the experiment and the modulation strength is of the same order of magnitude \cite{pengMidinfraredElectroopticModulation2017c}. The quantitative discrepancy can be due to the fact that additional unknown experimental factors e.g. disorder and detailed dielectric environment should be taken into account.

To conclude, we investigated theoretically the optical properties of few-layer phosphorene in the presence of a perpendicular electric field. By means of a self-consistent TB approach combined with the standard Kubo formula, we calculated the interband optical conductivity of biased few-layer phosphorene over a wide range of photon energies covering the infrared-to-visible frequency range.

We found that (i) the interband optical conductivity shows a large linear dichroism that can be modulated by the perpendicular electric field; (ii) the field-induced charge screening plays a significant role in modulating the optical conductivity and the linear dichroism; (iii) the conductivity spectrum shows distinct absorption peaks due to the strong quantum confinement along the out-of-plane direction and the field-induced forbidden-to-allowed transitions; (iv) the field modulation strength of the interband optical transmission oscillates as a function of the photon energy, with negative and positive values due to different interband optical transitions; (v) the field modulation of the linear dichroism becomes more pronounced with increasing number of phosphorene layers; and (vi) the modulation strength exhibits an asymmetry for opposite field directions due to the substrate-induced hole doping. We explained the optical absorption features in terms of the optical selection rules related to the parities (symmetries) of the electron and hole wave functions, which are numerically reflected in the momentum matrix elements of the interband optical transitions. We also found that due to the reduced lattice symmetry, the optical Hall effect such as the Faraday rotation is present in few-layer phosphorene even in the absence of an external magnetic field. The Faraday rotation is greatly influenced by the field-induced charge screening and is strongly dependent on the strength of perpendicular electric field and on the number of phosphorene layers. Our theoretical results are relevant for electrically tunable linear dichroism and Faraday rotation in few-layer phosphorene in the infrared-to-visible frequency range.

\section{Acknowledgments}

This work was financially supported by the Flemish Science Foundation (FWO-Vl) and by the FLAG-ERA project TRANS-2D-TMD.

\bibliographystyle{apsrev4-1}

%

\end{document}